\documentstyle[12pt]{article}

\def\({\c c}
\def\ii {\'\i}

\def\nl {\par \noindent }

\begin{document}

 \sloppy

\hoffset = -1truecm
\voffset = -2.5truecm  


\title{{\begin{flushright}
\nl {\normalsize SLAC-PUB-8252}
\nl {\normalsize September 1999}
\end{flushright}}
\vspace{0.4cm}
\large\bf CHIRAL BOSON THEORY ON THE LIGHT-FRONT \thanks{Research partially supported by the Department 
of Energy under contract DE-AC03-76SF00515.}} 
\vspace{2cm}

\author{
{\large \bf
Prem P. Srivastava\thanks{E-mail: (1) prem@lafexsu1.lafex.cbpf.br;   
(2) prem@slac.stanford.edu. On leave of absence from {\it Instituto de 
F\ii sica, UERJ- Universidade do Estado de Rio de Janeiro}, RJ, Brasil. } }\\
\it Stanford Linear Accelerator Center, Stanford University, Stanford,\\
{\it CA 94309, USA.} }

\baselineskip=18pt  %
\date{}

\maketitle

\vspace{0.7cm}

\begin{abstract}

The {\it front form} framework for describing the quantized theory of 
chiral boson is discussed. 
  It avoids the  conflict with the requirement of the 
principle of microcausality as is found in the conventional {\it instant form} 
treatment.  The  discussion of the Floreanini-Jackiw model and 
its modified version for describing the chiral boson  
becomes very transparent on the light-front. 

\end{abstract}
\vfill

\nl {\bf Keywords}:\quad{Chiral boson, Selfdual fields, 
Light-front quantization }

\vfill

\newpage

\nl 1-\quad There is a wide interest in the chiral bosons, 
also called self-dual  scalar 
fields \cite{schw}. They are, for example, among the basic ingredients in 
the formulation of heterotic  string theory 
 \cite {gross},   
in the description \cite{stone} 
of boundary excitations  of the quantum Hall state,  
in a number of two-dimensional statistical systems \cite{boya, west} 
which are related 
to the Coulomb-gas model, and in the context of W-gravities. 
Although apparently simple, the quantization of chiral bosons 
 presents intriguing and instructive features. 
The canonical quantization of Siegel's Lagrangian \cite{siegel} model 
requires 
an additional Wess-Zumino term to take care of the anomaly. The resulting
theory does not describe pure chiral boson but rather their coupling to
gravity \cite{imb}.  The model which employs the Lagrange multiplier field 
to impose the chiral constraint linearly \cite{sriv} 
has received criticism \cite{rivel}.  
An  improved version \cite{girotti} of it, however, is found 
equivalent  at the quantum level 
 to the much studied model of chiral boson,  
proposed \cite{flor} earlier by Floreanini and Jackiw (FJ).  
The rather extensive 
literature, which employs the conventional equal-time framework in  
 its study,  reveals its polemic character. It is thus 
 worthwhile to study its quantization on the light-front (LF) which, as 
shown below, does throw some new light on the problem. If we take  into
consideration the requirement of the principle of microcausality, the {\it front
form} framework is seen    to be more appropriate    
for discussing  the quantized theory 
of the chiral boson. We  will study in some detail the FJ model 
 modified by the introduction of an additional   parameter in it.

\vspace{0.5cm}
\nl 2- \quad Half a centuary ago,  Dirac \cite{dir}  discussed the unification, 
in a relativistic theory,  of 
the principles of the quantization and the special relativity theory which were
by then firmly established. 
The Light-Front (LF) quantization which studies the  relativistic 
quantum dynamics of  physical system on the hyperplanes  
: $x^{0}+x^{3}\equiv {\sqrt{2}}x^{+}=const.$,  called 
the {\it front form} theory, was also proposed there and some of 
its advantages pointed out.  
  The {\it instant form} or the 
conventional equal-time theory 
on the contrary uses the $x^{0}=const.$ hyperplanes.  
The LF coordinates $x^{\mu}: (x^{+},x^{-},x^{\perp} 
)$,  where $x^{\pm}=(x^{0}{\pm} x^{3}) 
/{\sqrt 2}=x_{\mp}$ and   $ x^{\perp} = 
(x^{1}, x^{2})=(-x_{1},-x_{2})$,   are convenient to use in the {\it front form}
theory.  
They are  {\it not related by a finite Lorentz transformation} 
to the coordinates $(x^{0}\equiv t,x^{1},x^{2},x^{3})$ 
 employed in the {\it instant form } theory.  The 
descriptions of the same physical content in a dynamical theory 
on the LF, which studies the evolution of the system in $x^{+}$ in place of
$x^{0}$, 
 may thus come out to be different from that given in the conventional 
treatment. This is   found, for example, to be the case  
  in the description of the 
spontaneous symmetry breaking (SSB) mechanism \cite{pre} 
and in the  studies \cite{pre1}  of some soluble 
two-dimensional gauge theory models,
where it was also demonstrated 
that LF quantization is very economical in displaying the
relevant degrees of freedom, leading  directly to the physical
Hilbert space.

  The interest in the {\it front form} theory 
 has been revived \cite{bro, ken, pre}   
 due to the difficulties encountered in the computation, 
in the conventional 
framework,  of the nonperturbative effects in the context of  QCD 
and  the problem of the relativistic bound states of  
fermions \cite{bro, ken} 
in the presence of the complicated vacuum. 
 LF variables have  found  applications in several contexts, 
for example, in 
the quantization of (super-) string theory and M-theory \cite{susskind1},   
in the nonabelian
bosonization \cite{wit} of the field theory of $N$ free Majorana fermions, in   
the study of the vacuum structures \cite{pre1} in the  
 Schwinger model (SM) and the 
Chiral SM  among many others.  The LF quantized QCD in covariant gauges has also 
 been studied \cite{pre3} recently 
 in the context of the Dyson-Wick perturbation theory,  
where it is shown that  the apparent  lack of manifest covariance 
usually encountered in such   calculations  becomes  tractable 
 thanks to the introduction of a useful construction of the LF spinor. 
 The applications in the context of the Bethe-Salpeter dynamics 
 have also been considered \cite{mit2,carbonel} recently.

\vspace{0.5cm}


\nl 3-\quad We will make the {\it convention} to regard
$x^{+}\equiv \tau$ as the 
LF-time coordinate while $x^{-}$ as the {\sl longitudinal 
spatial} coordinate. 
The temporal  evolution in $x^{0}$ or 
$x^{+}$ of the system is 
generated by the Hamiltonians which are different 
in the two {\it forms} of the theory.

Consider \cite{pre1} the invariant distance between two spacetime points 
: $ (x-y)^{2}=(x^{0}-y^{0})^{2}-(\vec x-\vec y)^2= 2 
(x^{+}-y^{+}) (x^{-}-y^{-}) - (x^{\perp}-y^{\perp})^{2}$. 
On an equal $x^{0}=y^{0}=const. $ hyperplane the points have  
spacelike separation  except for if  they 
are {\it coincident} when it becomes lightlike one.  
On the LF with $x^{+}=y^{+}=const.$ 
the distance becomes  {\it independent of}  $(x^{-}-y^{-})$ and 
the seperation is again spacelike; it becomes lightlike one 
when  $x^{\perp}=y^{\perp}$ but with the difference that 
now the points need {\it not}  
necessarily be coincident along the longitudinal direction. 
The LF field theory hence {\it need not necessarily be local} 
in $x^{-}$, even if the corresponding 
{\it instant form} theory is formulated as a local  one. 
For example, the commutator 
$\;[A(x^{+},x^{-},{x^{\perp}}),B(0,0,0^{\perp})]_{x^{+}=0}\;$ 
of two scalar observables would vanish on the grounds of
microcausality principle in relativistic field theory for 
$ x^{\perp}\ne 
0$ when  $x^{2}\vert_{x^{+}=0}$ is spacelike. 
Its value  would  hence be proportional to  $\,\delta^{2}(x^{\perp})\, $ 
and a finite number of its derivatives,  
implying locality only in $x^{\perp}$ but not necessarily so 
in $x^{-}$. Similar arguments in 
the {\it instant form} theory lead to the locality 
in all the three spatial coordinates. 
In view of the microcausality \cite{weif} both  of the commutators 
 $[A(x),B(0)]_{x^{+}=0}$ and 
$[A(x),B(0)]_{x^{0}=0}$ are nonvanishing    
only on the light-cone.

 An  important advantge of the {\it front form} theory 
pointed out by Dirac  is that here  seven out of the ten   
Poincar\'e generators are {\it kinematical} while there are only six such ones 
in the  the conventional theory.  
Moreover, based   on the general considerations it can be argued \cite{pre} 
  that the LF hyperplane is equally valid and appropriate as the conventional
equal-time one for the canonical quantization of relativisitic theory.

\vspace{0.5cm}


\nl 4-\quad  The massless two dimensional  free scalar   theory has, at       
 the      classical level, the chiral boson 
solutions satisfying  $\partial_{0}\phi=\pm \partial_{1}\phi $.    
 We would like to  construct the corresponding 
Lagrangian formulation which describes, in the quantized theory, the 
excitations of, say, a right-moving massless particle.

The FJ model   is  based on the 
 following {\it manifestly  non-covariant Lagrangian}
\begin{eqnarray}
{ \cal L}&= & (\partial_{0}\phi-\partial_{1}\phi)\partial_{1}\phi 
\nonumber \\
&= &
\frac{1}{2}{\eta}^{\mu\nu} \partial_{\mu}\phi \partial_{\nu}\phi -
\frac{1}{2} (\partial_{0}\phi-\partial_{1}\phi)^{2}.  
\end{eqnarray}
where $\phi$ is a real scalar field and 
$\eta^{00}=- \eta^{11}=1, \; \eta^{01}=\eta^{10}=0$.  
In the {\it instant form }   
frame work its canonical quantization  results in \cite{flor, costa}
  the following equal-time commutator 
\begin{equation}
\left[\phi(x^{0},x^{1}), \phi(x^{0}, y^{1})\right]= \frac{-i}{4}
\epsilon (x^{1}-y^{1}).  
\end{equation}
It is nonlocal and  nonvanishing for spacelike distances, e.g., it 
 {\it violates} the microcausality principle, 
contrary to what we encounter \cite {weif} normally 
in the   conventional  theory  framework.      
 These  objections are found  below to  {\it  disappear}   
if we  regard the theory under discussion 
as being considered in the {\it front form} framework. 

 The FJ Lagrangian (1) may, in fact, be   rewritten 
in terms of the LF coordinates
as  $\; {\cal L}= (\partial_{+}\phi-
\partial_{-}\phi)\;\partial_{-}\phi $.  We will consider instead the following 
modified form of the Lagrangian for describing the chiral boson 
\begin{eqnarray}
{ \cal L}&= & (\partial_{+}\phi-\frac{1}{\alpha}
\partial_{-}\phi)\;\partial_{-}\phi 
\nonumber \\
&= &
\frac{1}{2} {\eta}^{\mu\nu} \partial_{\mu}\phi \partial_{\nu}\phi - 
 \frac {1}{\alpha}(\partial_{-}\phi)^{2},  
\end{eqnarray}
where $\eta^{+-}=\eta^{-+}=1, \eta^{++}=\eta^{--}=0$, $\mu,\nu= \pm $, and  
$\alpha$ is a fixed parameter.  
The canonical momentum following from (3) 
is $\pi= \partial_{-}\phi$, which indicates 
that we are dealing with  a constrained dynamical system. The Dirac
procedure \cite{dir1} or the Faddeev and Jackiw method \cite{faddeev} may be
followed  to construct the  Hamiltonian framework 
which in its turn may  be  quantized 
canonically. The discussion in our case is straightforward and 
follows   closely 
 the one given in  refs. \cite{flor, costa} and we    
are  lead  to the following LF Hamiltonian 
\begin{equation}
{ H}^{lf}= \int d x^{-} \; \frac{1}{\alpha} (\partial_{-}\phi)^{2}.
\end{equation}
 The equal-$\tau$ commutator is derived  to be 
\begin{equation}
\left[\phi(\tau,x^{-}),\phi(\tau,y^{-})\right]= \frac{-i}{4}
 \epsilon(x^{-}-y^{-})
 \end{equation}
while $\pi$ gets eliminated  from the theory. 
The  LF commutator (5) is nonlocal in $x^{-}$ and  
  nonvanishing only on the light-cone. It  does  {\it not}  
   conflict with the microcausality principle  unlike 
the equal-time commutator (2) in the {\it instant form }theory.   
  The Heisenberg equation of motion for the field operator is found to be 
\begin{equation}
\partial_{+}\phi=\frac{1}{i}\left[\phi, H^{lf}\right]=
 \frac{1}{\alpha}\partial_{-}\phi  
\end{equation}
and   the Lagrange equation 
\begin{equation}
  \partial_{-}\left[\partial_{+}\phi - \frac{1}{\alpha}\partial_{-}\phi
 \right]=0.
 \end{equation}
is recovered.

The commutator (5) can be realized in momentum space through the following 
Fourier transform of the field
\begin{equation} 
\phi(\tau,x^{-})=\frac{1}{\sqrt{2\pi}}\!
\int dk^{+}\;\!\frac{\theta(k^{+})}{\sqrt{2k^{+}}}
\left[a(\tau, k^{+}) e^{-ik^{+}x^{-}}+ 
{a^{\dag}}(\tau, k^{+}) e^{ik^{+}x^{-}}\right],  
\end{equation}
if  the creation and annihilation operators $a^{\dag}$ and $a$ 
are  assumed to satisfy the equal-$\tau$ 
canonical commutation relations, with the nonvanishing one given by 
 $\;\left[a(\tau, k^{+}), 
{a^{\dag}}(\tau, p^{+})\right]= \delta(k^{+}-p^{+})$.
On using the equation of motion (6) we derive easily   
\begin{equation}
a(x^{+}, k^{+})= e^{-ik^{-}x^{+}} a(k^{+}), \qquad 
a^{\dag}(x^{+}, k^{+})= e^{ik^{-}x^{+}} a^{\dag}(k^{+}),
\end{equation} 
where 
\begin{equation}
k^{-}=\frac{1}{\alpha} k^{+}, \qquad \qquad \mbox{\rm implying}
\qquad\qquad 2k^{+}k^{-}=\frac{2}{\alpha}(k^{+})^{2}.  
\end{equation}
    The Fourier transform of the field then   assumes  the form 
\begin{equation} 
\phi(x^{+},x^{-})=\frac{1}{\sqrt{2\pi}}\!\int dk^{+}\;
\!\frac{\theta(k^{+})}{\sqrt{2k^{+}}}
\left[a( k^{+}) e^{-ik \cdot x}+ 
{a^{\dag}}( k^{+}) e^{ik \cdot x}\right]. 
\end{equation}
where $k\cdot x\equiv k^{-}x^{+}+ k^{+}x^{-}
= k^{+} (x^{-}+x^{+}/\alpha)$ and 
the nonvanishing commutator satisfies  $\;\left[a( k^{+}), 
{a^{\dag}}( p^{+})\right]= \delta(k^{+}-p^{+})$.

We recall now that on the LF the dispersion relation associated with the free
massive particle is  $2p^{+}p^{-}=(p^{\perp}p^{\perp}+m^{2})>0$. 
It has  no  square root,  like 
in the conventional case $\;p^{0}=\pm \sqrt{{\vec p}^{2}+m^{2}}$. 
The  signs   
of $p^{+}$ and $p^{-}$ are thus {\it correlated}  in view of 
 $\;p^{+}p^{-}>0\,$. For  massless particles   
the correlation   ceases to exist
at the point  $p^{\perp}\to 0$ when  $2 p^{+}p^{-}=p^{\perp}p^{\perp}\to 0$. 
 On the other hand, for finite values of $\alpha$,  
 the  dispersion relation (10) obtained above is different 
from that of a free massless particle.   
Only  in the limit when  $|\alpha| \to \infty\;$ 
does $2k^{+}k^{-}\to 0$ and, according to 
(11), $\phi\to \phi_{R}(x^{-})$, which describes  a
right (moving) chiral boson.

 Consider next the components of 
  the classical canonical energy-momentum tensor $T^{\mu \nu}$. We find  
\begin{eqnarray}
T^{+-}=-T^{-+}= \frac{1}{\alpha} T^{++}&=& 
\frac{1}{\alpha} (\partial_{-}\phi)^{2},\nonumber\\
T^{--}&= &(\partial_{+}\phi)^{2}-\frac{2}{\alpha}(\partial_{+}\phi)
(\partial_{-}\phi). 
\end{eqnarray}
They obey the     on shell conservation equations  
\begin{equation}
 \partial_{\mu}T^{\mu \pm}= 2 (\partial_{\mp}\phi)\,
 \partial_{-}\left[\partial_{+}\phi - \frac{1}{\alpha}\partial_{-}\phi
 \right]=0
\end{equation} 
as    may be easily
checked. We may thus   define,  
if the surface integrals can be ignored,  
 the following conserved translation generators  
\begin{equation}
P^{+}= \int d x^{-} :T^{++}:\;= \int d x^{-} :(\partial_{-}\phi)^{2}:\;=
\int dk^{+} \theta(k^{+}) \; N(k^{+})\; (k^{+}) 
\end{equation}
and 
\begin{equation}
P^{-}\equiv H^{lf}= \int d x^{-} :T^{+-}:\;= \frac{1}{\alpha}\; P^{+}, 
\qquad \mbox{\rm implying}
\qquad 2P^{+}P^{-}=\frac{2}{\alpha}(P^{+})^{2}.
\end{equation}
Here  $N(k^{+})={a^{\dag}}( k^{+})a( k^{+})\;$ is the number operator and 
$\,:\;\; :\,$ indicates the normal ordering.

From (13) and in  virtue of $ \;(T^{+-}+T^{-+})=0$, following from (12),     
we  derive 
\begin{equation}
\partial_{+}\left[x^{-} T^{++}+ x^{+} T^{+-}\right]+
\partial_{-}\left[x^{-} T^{-+}+ x^{+} T^{--}\right]= 0,  
\end{equation}
which is valid on shell.  
It hence allows us to define the following   conserved symmetry generator  
\begin{equation}
M=  x^{+} P^{-} + \int dx^{-}\; x^{-} T^{++}.
\end{equation}
The generators $M, P^{+}, P^{-}$ are shown  to form 
a closed algebra : \quad  
  $\left[ M, P^{+}\right]
=-iP^{+}$,   $\; \left[ M, P^{-}\right]=-iP^{-}$, and 
$\;  \left[ P^{+}, P^{-}\right]= 0$. The operator  $M$ thus 
generates the scale (boost) transformations on $P^{\pm}$ 
by the same amount which leaves $P^{+}/P^{-}$ invariant.   
The {\sl mass operator} $\,2P^{+}P^{-}$, however, does get  scaled and 
is {\it not} invariant  under the transformations generated by $M$. 
The Lagrange equation is easily 
 shown to be form invariant under the 
infinitesimal symmetry transformation  
$\; \phi\to \phi+ \epsilon ( x^{-}+ x^{+}/\alpha )\partial_{-}\phi\;$   
generated by $M$. The operator $M$ clearly  
resembles  the (kinematical) 
  Lorentz boost generator, {\it viz},  
$\, M^{+-}\equiv - x^{+} P^{-}+ \int dx^{-}\;
x^{-} T^{++}\,$ which, as seen from (12) and  (13),  however,       
is  {\it not conserved}   in the manifestly  noncovariant model   
  under consideration.  


 In the limit  when $|\alpha| \to \infty$  we 
find $\;\phi \to {\phi_{R}}(x^{-})\;$ while $H^{lf}\to 0$,  like in the case of the 
 LF Hamiltonian of the  free  massless scalar theory. 
The field commutator of $\phi_{R}$ 
is found to be : $\; \left[{\phi_{R}}(x^{-}),{\phi_{R}}(y^{-})\right]= 
-i \epsilon(x^{-}-y^{-})/4$. The limiting case is thus shown  
to describe  a right (moving) chiral boson theory with  the Lagrangian density 
 given in (3).   

 An alternative form of the Lagrangian density is also possible. 
  We recall that  in the  quantization of  gauge theory in $3+1$ dimensions 
    it is  found useful  to  introduce  an auxiliary Nakanishi-Lautrup 
 field $B(x)$ of canonical mass dimension two and   
  add to the Lagrangian density  $\;(B\partial_{\mu}A^{\mu}
 +\alpha B^{2}  )\;$ as the gauge-fixing term. 
In the two dimensional theory under consideration 
 it is also  possible to follow 
this procedure,     
since the corresponding   $B(x)$  field here 
carries the canonical mass dimension one.  The 
discussion parallel to the one given 
above may    be  based \cite{barcel} equally well 
 on    the following  {\it front form } Lagrangian density
\begin{equation}
 {\cal L}=\frac{1}{2} {\eta}^{\mu\nu} \partial_{\mu}\phi \partial_{\nu}\phi 
 +{\sqrt 2} B(x)(\partial_{-}\phi) + \frac{\alpha}{2} B(x)^{2}. 
\end{equation}
If we  eliminate in it   the auxiliary field $B$ by 
using its equation of motion it leads back to (3).  
  The conclusions following from 
 the one or the other are the same.

 We  make only brief comments on the  other models. 
   Siegel's \cite{siegel} theory  which employs
\begin{equation}
{ \cal L}= \frac{1}{2}{\eta}^{\mu\nu} \partial_{\mu}\phi \partial_{\nu}\phi +
B(x) (\partial_{0}\phi-\partial_{1}\phi)^{2} 
\end{equation}
is afflicted by anomaly which is to be eliminated by the addition of 
a Wess-Zumino term. The resulting theory does not describe \cite{imb}
pure  chiral bosons since they are coupled to the gravity. 
 In this model the 
auxiliary field carries vanishing canonical dimension and, for 
example, a $B^{2}$ term cannot be added to it without introducing the 
dimensionful parameters. The  model  based on the idea of implementing the 
chiral constraint through a linear constraint \cite{sriv, barcel},  
\begin{equation}
{\cal L}=\frac{1}{2} {\eta}^{\mu\nu} \partial_{\mu}\phi \partial_{\nu}\phi   
 +B_{\mu}({\eta}^{\mu\nu}-\epsilon^{\mu\nu})
\partial_{\nu}\phi,
\end{equation}
where $B_{\mu}$ is Lagrange multiplier field,    
does not seem to exhibit physical excitations \cite{rivel}. We note that the 
 field  $B_{\mu}$ carries dimension one and that this is the usual procedure 
in the classical theory which, however,    breaks down 
in the quantized theory. 

\vspace{0.5cm}
\nl {\bf Conclusions}
\vspace{0.3cm}

  The {\it front form} quantized theory of chiral boson 
is straightforward to construct.  It may be  based on the modified FJ model 
as described by the Lagrangian density  (3) or in its alternative 
form (18). The discussion becomes  transparent on the LF and  it  
also avoids the  conflict with the requirement of the principle of 
microcausality,   in contrast to what  found in the corresponding 
{\it instant form } theory.   

\vspace{0.7cm}

\nl {\large \bf Acknowledgements}
\vspace{0.4cm}

The author acknowledges with thanks the helpful 
comments from Sidney Drell, Stan Brodsky, and 
Michael Peskin.  
The hospitality offered to him at the SLAC and a financial grant 
of Proci\^encia 
program of the UERJ, Rio de Janeiro, Brasil, are gratefully acknowledged.

\vspace{1cm}


\end{document}